\documentclass{PoS}

\title{Critical Opalescence: \\
An Optical Signature for a QCD Critical Point}

\ShortTitle{Critical Opalescence} 

\author{
	Tam\'as Cs\"org\H{o}
\thanks{Supported by OTKA grants T49466 and NK 73143, and by a HAESF Senior Leaders and Scholars Award.}\\
        Department of Physics, Harvard University, 17 Oxford St, Cambridge, MA 02138, USA\\
        MTA KFKI RMKI, H-1525 Budapest 114, POBox 49, Hungary \\
        E-mail: \email{csorgo@physics.harvard.edu}}

\abstract{
Four possible scenarios are considered  
for a transition from a quark-gluon matter to hadronic matter, 
and their corresponding correlation signatures are discussed.
Four criteria are highlighted for  
a definitive experimental search for a QCD critical point.
An old-new experimental measure, the optical opacity (or its inverse
the nuclear attenuation length) is determined, 
in terms of a combination of nuclear suppression factors and a measurement 
of the relevant fireball length scales. 
Length scale estimates using either the Hanbury Brown -- Twiss radii 
or that of the initial nuclear geometry for measurements of optical opacity with respect 
to the reaction plane yield, somewhat surprizingly, nearly the same nuclear attenuation
lenght in 0-5 \% most central 200 GeV Au+Au collisions,  corresponding to 2.9 $\pm$ 0.3 fm. 
The necessity and the possibility of measuring critical exponents is also discussed  
in the context of determination of the universality class of the QCD critical point.
Critical opalescence is proposed to locate such a critical point 
on the QCD phase diagram, corresponding to a maximum of optical opacity 
in heavy ion experiments.
}

\FullConference{5th International Workshop on Critical Point and Onset of Deconfinement - CPOD 2009,\\
		 June 08 - 12 2009\\
		 Brookhaven National Laboratory, Long Island, New York, USA}

\begin{document}

\section{Introduction}
One of the major goals of high energy heavy ion physics is the exploration of the
phase diagram of strongly interacting hot and dense matter. The exploration of this diagram 
is of fundamental interest. It allows for exploring the properties of
early Universe a few microseconds after the Big Bang, but also
to understand the collective properties of a non-Abelian gauge theory,
Quantum Chromo Dynamics (QCD), the theory of strong interactions,
and to explore the dynamics of color confinement. 

A schematic, theoretical phase diagram of QCD is shown in Fig.~\ref{fig:QCD-phase-diagram}.
At high temperatures and / or net baryon densities one expects, based on asymptotic freedom,
that a gas of weakly interacting, nearly massless quarks and gluons exists. A popular model
of this state was formulated in terms of an ideal Bose-gas   equation of state for gluons
and an ideal Fermi gas of quarks, and a bag constant was introduced to model confinement.
This simple bag model incorporates a first order confinement-deconfinement phase transition.
 
At the end of the line of the first order QCD phase transitions, a critical end point (CEP) is supposed
to exist, where the phase transition becomes of second order. 
Locating and characterizing this point on the QCD phase
diagram in a definitive manner could be of a great significance: 
if such a point exists, it separates a line of first order deconfinement/chiral symmetry restoration phase transitions 
from a smooth cross-over region. 
Cross-over transition means that in a narrow temperature and density region the degrees of freedom change
quickly but, in a strict thermodynamical sense, there are no phase transitions any more and in this region 
different physical
quantities are extremal at different values of the temperature as discussed e.g. in ref.
~\cite{Fodor:2004nz}.

\section{Four Criteria for  Definitive Measurements of a QCD Critical Point} 
\begin{figure}[tb]
\begin{center}
\includegraphics[width=0.7\textwidth]{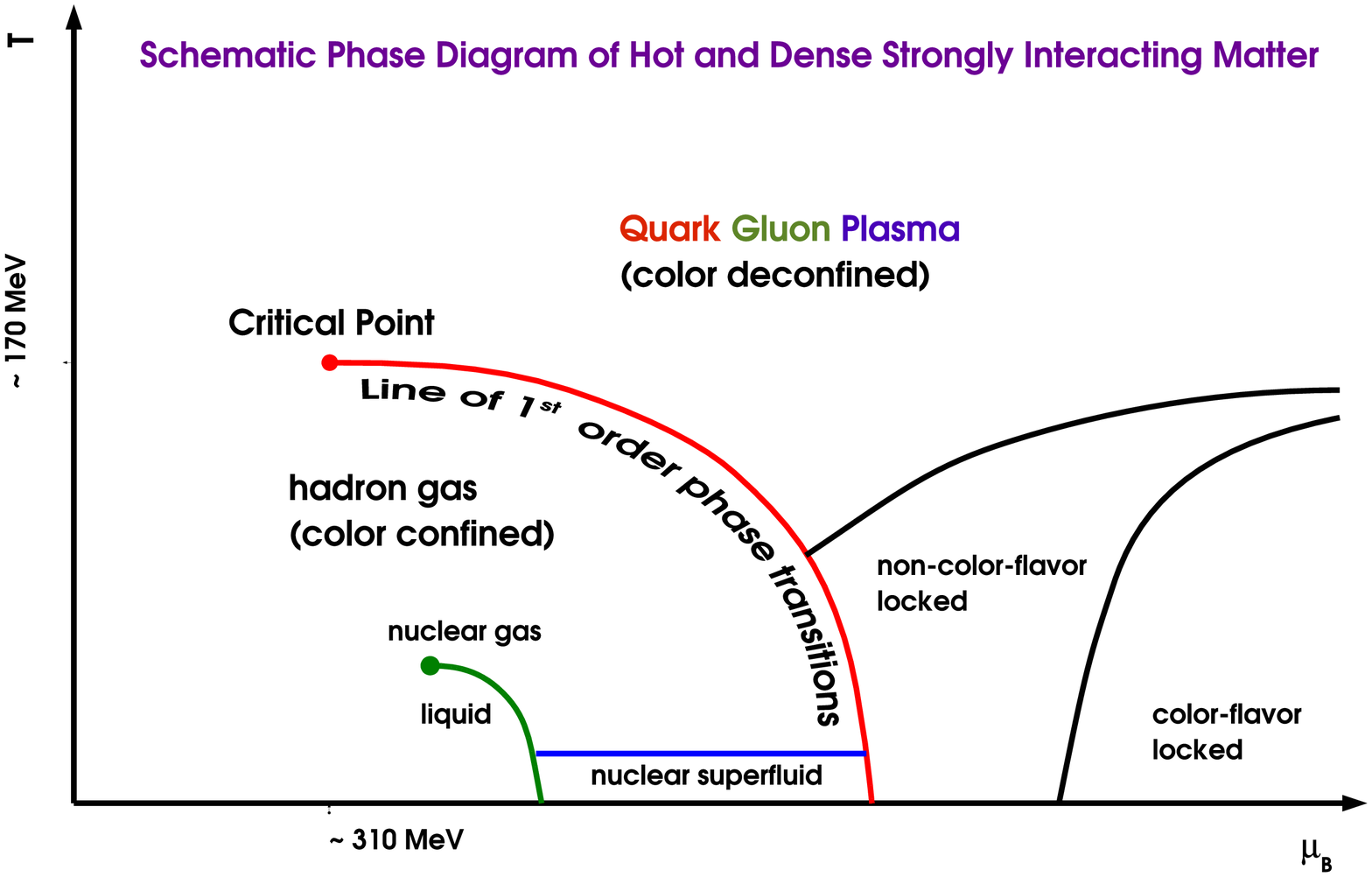}
\end{center}
\caption{%
\label{fig:QCD-phase-diagram}
Schematic phase diagram of QCD based on ref.~\cite{Alford:2007xm}, reproduced from 
ref.~\cite{Csorgo:2009gb}. Two critical points are shown: at lower values of
temperature, the nuclear liquid-gas phase transition line ends in a critical point,
while at higher temperatures the so called ``QCD critical point" is shown, 
as  the end of the line of first order deconfinement/chiral symmetry restoration phase transitions. 
}
\end{figure}

Although the experimental identification of the QCD critical end point of the line 
of the first order phase transitions is of great significance, this task is not an obvious one.
Although it is not shown in the illustration of Fig.~\ref{fig:QCD-phase-diagram}, recent results suggests 
that more than one such critical points of QCD are theoretically possible~\cite{Bowman:2008kc}. 
Alternatively, it is possible that no critical point exists in QCD,
as discussed in ref.~\cite{deForcrand:2008rp}: an empty phase diagram cannot be excluded at present. 

Ref.~\cite{Fodor:2004nz}
predicted  $ T_E = 164 \pm 2$ MeV and $\mu_E = 360 \pm 40 $ MeV
for the location of the end point of the line of first order QCD deconfinement
phase transitions, however, the errors are statistical only and different theoretical
approximations yield different values for this location. 

An alternative possibility is that a critical point in fact exists, but it 
it may be located in a region not accessible by experiments at CERN SPS or BNL RHIC.
However, an experimental search for a critical point in heavy ion collisions
is not a mission impossible:
A power-law distribution of intermediate mass fragments  was found in
intermediate energy in heavy ion reactions, and related to the  
critical point of a  nuclear liquid-gas phase transition~\cite{Mahi:1988hj}, \cite{Srivastava:2002xx}.
This nuclear liquid-gas critical point is also shown schematically in Fig.~\ref{fig:QCD-phase-diagram}.
Note, that  chiral symmetry restoration and deconfinement phase transition 
are in principle different kind of phase transitions, they may have two different
lines of first order phase transitions in the $(\mu_B, T)$ plane,
that may have two different critical end points,
however their critical temperatures may have
two closely, dynamically  related values, which are numerically similar~\cite{Braun:2009gm}.
Note furthermore,  that color confinement is a sound property of all the observed hadronic states, 
which is not reflected on the schematic phase diagram of Fig.~\ref{fig:QCD-phase-diagram},
that allows for color deconfined states at $T < T_c$ temperatures, 
with thermally suppressed, small but non-zero probability,
due to the cross-over 
at small chemical potential.

For a definitive, quality search for the QCD critical point(s), 
I think that the following four criteria have to be satisfied: 
we have to {\it identify}, {\it locate}, 
{\it characterize} and {\it cross-check} the properties of QCD Critical End Point(s), 
by answering the questions outlined below. 

\begin{description}
\item{\underline{Identify:}}
What is the type of phase transition -- is it chiral symmetry restoration, or is it
the phase transition that leads to quark (de)confinement, 
or do these a priori  different type of phase transitions coincide 
a posteriori~\cite{Braun:2009gm}? {\it What are the experimental order parameters}
that are measurable not only in lattice QCD but also in laboratory experiments of heavy ion physics? 
Do we  find experimental indication for the creation of quarkionic matter~\cite{McLerran:2007qj}?

\item{\underline{Locate:}}
How many critical points exist on the accessible part of the phase diagram? 
At what temperatures and net baryon densities is (are) the CEP(s) located - 
what are the values of $(T_E,\mu_{E})$?
How can we reach this(these) point(s), 
i.e. what kind of collisions, at what centralities and what $\sqrt{s_{NN}}$
produce critical phenomena?

\item{\underline{Characterize:}}
In order to be able to relate the  properties of a QCD Critical End Point to other critical points of 
statistical / solid state physics and physical chemistry, e.g. that of water,
we should characterize and quantify the QCD Critical End Point(s) using the standard language
of statistical mechanics. To make such comparisons possible, we should attempt to measure the  
critical exponents,  and the universality class of the QCD critical point(s).

\item{\underline{Cross-check and control:}}
How significant is the determination of the critical exponents and the universality class?
Can we measure more than the minimally necessary two critical exponents?
Do the additional critical exponents satisfy the constraints 
given by the general theory of second order phase transitions, as discussed
in ref.~\cite{Csorgo:2009gb}? If a characteristic power-law behavior is 
found experimentally, we should cross-check if this  
behaviour is really due to a QCD CEP, or
due to e.g. anomalous diffusion~\cite{Csanad:2007fr}, fractal structures or other effects. 
\end{description}

\section{Four possible forms of QCD Phase Transitions and their correlation signatures}
For simplicity, let us consider the case when the chiral symmetry restoration and the deconfinement
phase transitions coincide~\cite{Braun:2009gm}.
Let us then consider, what kind of  the deconfinement/re-confinement type of phase transitions
are logically possible.  
In a heavy ion collision, confinement can be realized either in a thermal equilibrium or in a non-equilibrium manner. 
Thermal equilibrium states may pass, as a function of time, through a line  of phase transitions or through a cross-over region.
Phase transitions can be either with latent heat (first order phase transitions) or without latent heat
(second order phase transitions).
Thus we obtain four possibilities and four different kind of characteristic  correlation signatures,
see ref.~ \cite{Csorgo:2007iv} for details.

\underline{i) (Strong) first order phase transitions:}
	Several groups calculated the Bose-Einstein or HBT correlation function
	from hydrodynamical models assuming a first order phase transition in equilibrium,
	with a significant amount of latent heat. This is reflected in 
	long life-times and large widths of the time distribution of particle
	production. The corresponding picture is  that of
	a slowly burning cylinder of QGP~\cite{Rischke:1996em}. 
	This physics is signaled as $R_{out} \gg R_{side}$,
	regardless of the exact details of the 
	calculation~\cite{Pratt:1984su,Pratt:1986cc,Hama:1987xv,Bertsch:1988db,Bertsch:1989vn,Rischke:1996em}.
	Experimentally, $R_{out} \approx R_{side}$ 
	both at CERN SPS ~\cite{NA44,NA49} and at RHIC~\cite{STARout,PHENIXout,PHOBOSout}.
 	Thus a slowly burning, predominantly longitudinally
	expanding  QGP ``log" is excluded by correlation measurements in $A$+$B$ reactions
	both at CERN/SPS and at RHIC.  Three alternatives remain: either a second order or a cross-over 
	or a non-equilibrium phase transition.

\underline{ii) A second order deconfinement phase transition}
	and its signature in Bose-Einstein/HBT correlations 
	was discussed recently  in ref.~\cite{Csorgo:2005it}. 
Such second order phase transitions are characterized by 
critical exponents. One of these,
traditionally denoted by $\eta$, characterizes the tail of the
spatial correlations of the order parameter. 
This exponent was shown to be
measurable with the help of the two-particle correlation function
~\cite{Csorgo:2005it}, where it was shown also that $\eta = \alpha$, where
$\alpha$ stands for the L\'evy index of stability of the
correlation function itself, i.e., $C(q) = 1 + \lambda \exp[ - (q
R)^\alpha ]$.  The excitation function of the {\it
shape parameter} $\alpha$ has unfortunately
not yet been studied experimentally, although such a measurement
may lead to the experimental discovery of the location of the  
QCD Critical Point at the minimum of this exponent. 

\underline{iii) Cross-over transitions:}
According to recent lattice QCD calculations,
 if the temperature is increased at zero, or nearly zero chemical potentials, 
the transition from confined to deconfined matter 
is a cross-over~\cite{Fodor-Katz-crossover},
and various observables yield different estimates for the critical
temperature itself: the peak of the renormalized chiral susceptibility predicts $T_c$=151(3) MeV,
whereas critical temperatures based on the strange quark number
susceptibility, and Polyakov loops, result in values higher than
this by $24(4)$ MeV and $25(4)$ MeV, respectively.  
This scenario can be signaled by emission of hadrons from a region 
above the critical temperature,
$T > T_c$, or, by finding deconfined quarks or gluons at
temperatures $T < T_c$.  Indeed, emission of hadrons from a small but superheated 
region, with $T > T_c \simeq 176 \pm 7$ MeV~\cite{Aoki:2006br,Fodor-Katz-CEP}, has been suggested by
Buda-Lund hydro model fits to single particle spectra and
two-pion HBT radii in Au+Au collisions at
RHIC~\cite{BL-indication}.

\underline{iv) A non-equilibrium transition: hadron flash from a supercooled QGP}.
In relativistic heavy ion collisions  at CERN SPS and at RHIC
the expansion time-scales are $\sim 10 $ fm/$c$, which are
short as compared to the $\sim 100$ fm/$c$ nucleation times of hadronic bubbles
in a first order phase transition from  a QGP to a hadron gas. 
Ref.~\cite{Csorgo:1994dd} suggested, that a rapidly expanding QGP fluid
 might strongly supercool in Au+Au collisions at RHIC and then rehadronize 
suddenly while emitting particles in a flash.  Such a sudden recombination from quarks to hadrons
has been considered at CERN SPS energies as well,
 as the mechanism for hadron formation in the
ALCOR model~\cite{ALCOR}. Other realizations of sudden hadron
formation and freeze-out were used to explain the observed
scaling properties of elliptic flow~\cite{v2-scaling} with the
number of constituent quarks. Ref. ~\cite{Csorgo:1994dd} 
suggested that the comparable magnitude of
$R_{out} \approx R_{side} \approx R_{long}$ could be a signature
of a non-equilibrium flash of hadrons (pion-flash) from a deeply supercooled QGP
phase~\cite{Csorgo:1994dd}. 
Note, that the thermodynamic considerations in
~\cite{Csorgo:1994dd} also allow, 
that from a supercooled QGP a super-heated hadron gas is formed
in about 50 \% of the parameter space.
The calculations of refs.\cite{Csorgo:1994dd,BL-indication}, even currently, 
are not in a clear disagreement with RHIC data on Au+Au
collisions (with the possible exception of an in-medium $\eta^\prime$ 
mass modification signal reported in ref.~\cite{Vertesi:2009ca}).
There are other more exotic possibilities for new phases, like 
color superconducting quark matter,
or quarkyonic matter, but these are not discussed here as
we focus on the search for the QCD critical point.

\section{Locating the Critical Point by Critical Opalescence} 

Before we attempt to locate and characterize a critical point in QCD, it can be
useful to consider, how a critical point is located in a more conventional scenario, 
for example in  statistical physics and in chemical experiments on second order phase transitions 
involving ordinary gases and fluids.  
The presentation in this section follows the line of ref.~\cite{Csorgo:2009gb}, 
where the above point of view lead the present author to a new, experimental definition of 
opacity in high energy heavy ion physics, hereafter referred to as ``optical opacity". 

\begin{figure}[tbp]
\begin{center}
\includegraphics[width=1.0\textwidth]{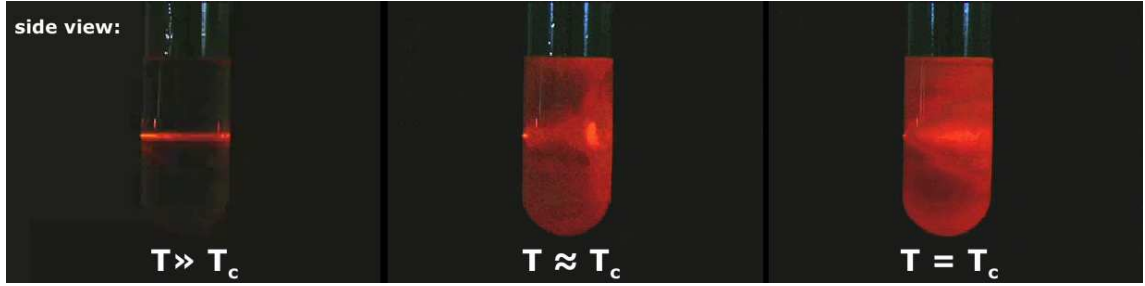}
\end{center}
\vspace{-0.5cm}
\caption{
 Critical opalescence is a smoking gun signature of a second order phase transition in
statitical physics and physical chemistry. A laser beam shining through
a test tube becomes more and more scattered and the fluid becomes more and more opaque, as the
critical point is approached.  These pictures are snapshots from 
a video of ref.~\cite{CEP:videos}. See also ref.~\cite{Csorgo:2009gb} for their first discussion 
in the context of search for a QCD critical point.
}
\label{fig:testtube-opacity}
\end{figure}

Critical opalescence is apparently a  smoking gun signature of a second order phase transition in
statitical physics and physical chemistry. Fig.~\ref{fig:testtube-opacity} indicates that a laser beam shining through
a test tube becomes more and more scattered and the fluid becomes maximally opaque 
at the critical point, where density fluctuations are present on all possible
lengthscales, including that of the wave-lenght of the penetrating beam.

In optics, opacity $\kappa$ is defined as the rate of absorption or extinction, at which the intensity
$I$ of a beam of radiation is absorbed or scattered per unit distance along a ray of propagation:
\begin{equation}
\frac{\partial I}{\partial x} = - \kappa I
\end{equation}
More generally the opacity $\kappa$ may depend on the wavelength (or momentum) of the radiation, as well as on the
density, temperature and composition of the medium. For example, air has nearly zero opacity for visible light,
and for radio waves, but it is almost completely opaque for gamma and X rays, and in most of the infrared
part of the spectrum.  The solution of the above differential equation for constant values of opacity is
\begin{equation}
I = I_0 \exp( - \kappa x)  = I_0 \exp(- x/\lambda). \label{e:opaque}
\end{equation}
In chemistry, this relation  is also observed and known as the Beer-Lambert law,
but it also appears in describing the change of intensity of an energetic $\gamma$ beam
passing through a lead target - simply if opacity  $\kappa $ is momentum and position independent, 
then the intensity of radiation decreases exponentially while it penetrates certain materials.
 
In the above equation $\lambda = 1/\kappa$ is the so called attenuation length or penetration depth, defined as the distance
where the detected radiation $I$ after passing through a material of width $\lambda$ falls 
to $1/e ~\approx 36.7 \%$ of its incoming intensity $I_0$.
Thus a quantitative signature of critical opalescence is a minimum of the attenuation length $\lambda$ or
a maximum of opacity $\kappa$ on many wavelengths simultaneously.
Measuring opacity as a function of pressure and temperature, the critical point can be located on
the phase diagram.

Let us follow ref.~\cite{Csorgo:2009gb} in rewriting the defining relation of optical opacity 
in a manner that has a straightforward application to heavy ion physics.
In heavy ion physics, the nuclear modification factor $R_{AA}$ can be defined as
\begin{equation}
R_{AA}= \frac{I(\mbox{\rm transmitted})}{I(\mbox{\rm generated})}
 = \frac{I(\mbox{\rm measured})}{I(\mbox{\rm expected})} = \frac{I}{I_0},
\end{equation}
where the measured yield is defined as
\begin{equation}
I(\mbox{\rm measured}) = \frac{1}{N_{event}^{AA}}\frac{d^2N_{AA}}{dy dp_t},
\end{equation}
while for point-like processes like production of high transverse momentum jets the generated
rate is the same yield measured in proton-proton reactions scaled up by $\langle N_{coll}\rangle$,
the number of binary initial nucleon-nucleon collisions, given by a Glauber model calculation:
\begin{equation}
I(\mbox{\rm expected}) = \frac{\langle N_{{coll}}\rangle}{\sigma^{NN}_{{inel}}}\frac{d^2\sigma_{NN}}{dy dp_t}.
\end{equation}

It is important to note that in high energy heavy ion reactions in the $p_t > 4$ GeV
kinematic range we are measuring not tiny absorbtions, but quite significant values, so
the thickness of the absorber or the fraction of the absorbed intensity cannot be considered
small, $1  - R_{AA} \approx 0.8$. Hence   we have to solve eq. (\ref{e:opaque}) to get
a more precise definition, keeping in mind the interpretation of the nuclear absorbtion factor
as $R_{AA}=I/I_0$. 
The nuclear modification factor $R_{AA}$ includes a mixture of hot and cold nuclear matter
effects, its most apparent shortcoming is that it measures the change in intensity but does not specify
the distance over which the change of the yield happened.
Opacity as defined in optics, on the other hand, compares the rate
of change of intensities per unit distance.
Thus a proper opacity  definition should include  both the nuclear modification factor
and the characteristic length scale of the attenuation.

A possible approach was proposed in ref.~\cite{Csorgo:2009gb} to measure
the source sizes utilizing Gaussian and L\'evy fits to two-particle Bose-Einstein correlation functions.
These radii are frequently referred to as Hanbury Brown - Twiss or
HBT radii to honor the two radio astronomers, who invented a similar technique for
photons to measure the angular diameter of main sequence stars.

It is well known that the size of nuclei change on the average as $A^{1/3}$, which in a collision
would then be modified to an average $N_{{part}}^{1/3}$
dependence.  PHENIX published the freeze-out volume (more precisely, the size of the region of homogeneity) of
charged pion emission in 200 GeV Au+Au
collisions using Bose-Einstein or HBT techniques as a function of centrality~\cite{Adler:2004rq}.
The Gaussian HBT radii were found to scale as
$R_i = p_0 + p_1 N_{{part}}^{1/3}$, where $p_0 \approx 0.83 \pm 0.25 $ fm and $p_1 \approx 0.54 \pm
0.05 $ fm and $i$ = (long, out, side) stands for the three principal directions with
respect to the beam and the mean transverse momentum of the particle pair.
Note that ref. ~\cite{Adler:2004rq}
fitted the  centrality dependence of each
HBT radius components separately, but 
in the different directions these radii were found, within one standard deviation,
 to have the same values~\cite{Adler:2004rq}. 
Thus their average value was quoted above.
The corresponding HBT volume as well
as the initial volume scales with $N_{part}^{1/3}$, i.e. a posteriori the initial static  volume
and the dynamically generated HBT source sizes are proportional.
Expansion effects may introduce absolute calibration errors on the opacity $\kappa$
that is difficult to control. However, the relative error on the centrality dependence of the
opacities $\kappa$ and in particular its minimum or maximum structure  is not affected
by these expansion effects, as both the initial and the final volumes scale proportional
to $N_{part}^{1/3}$.

Using an average value of the HBT radius as $R_{HBT} = (R_{{out}}+ R_{{side}}+ R_{{long}})/3$
and the nuclear modification factors of neutral pions measured by PHENIX it was possible
to evaluate the opacity and the attenuation lengths in $\sqrt{s_{NN}}=200$ GeV Au+Au collisions, as
a function of centrality. 
For arbitrary thickness and nuclear modification factor, the
optical opacity (and its inverse, the attenuation length) was determined~\cite{Csorgo:2009gb} as
\begin{equation}
\kappa = - \frac{ \ln(R_{AA})}{R_{HBT}}, \label{e:kappa}
\qquad \qquad \lambda = - \frac{ R_{HBT} } { \ln(R_{AA}) } . \label{e:lambda}
\end{equation}

 A simple test of such opacity measurement was reported in
ref.~\cite{Csorgo:2009gb}. 
The results are shown in Table~\ref{table:opacity}.  One finds that changing centrality
from 0-5 \% to 50 - 60\% , the opacity is decreased substantially, by about a factor of 2 (a 4 $\sigma$ effect). 
Thus nearly half of  the increased  nuclear suppression originates from the increased opacity of the
hot and dense, strongly interacting matter created in more central collisions,
while the rest arises due to an increased volume of matter produced in these collisions.

The attenuation lengths $\lambda$-s vary as a function of centrality in the range of 3-8 fm,
for neutral pions with $p_t = 4.75$ GeV/c, periferal events correspond to less stopping and
large attenuation lengths while for central collisions the attenuation lenghts decrease,
and the matter becomes more and more opague.

A similar approach to opacity determination is to constrain the relevant length-scales by measuring $R_{AA}$ relative to the 
reaction plane of the collision, as has been recently reported by PHENIX in ref.~\cite{Afanasiev:2009iv}.
In this paper, nuclear modification factors were plotted as a function of distances calculated from
hard sphere or from RMS approximations, and an exponential like suppression pattern,
just like the form of eq. (\ref{e:opaque}) was observed.
However, the attenuation lengths and optical opacities were not evaluated from
the available information in ref.~\cite{Afanasiev:2009iv}.  
In this approach, one  does not provide  a direct measurement of 
the spatial scales,  instead one relies on simple theoretical models of the initial
geometry, and the relevant lengthscales are evaluated in a hard sphere or in an rms approximation.  
In fact, the distance that is covered by penetrating probes in the medium is not a static distance, but a length scale
of the dynamical, expanding source that is covered by the jets during their punch through.

\begin{table}[ht]
\centering
\begin{tabular}{l c c c c c}
\hline\hline
Centrality &  0-5 \% &  20-30 \% & 30-40 \% & 40-50 \% & 50-60 \%\\ [0.2ex]
\hline
Opacity $\kappa$ (fm$^{-1}$) & 0.35  $\pm$ 0.04 & 0.27 $\pm$ 0.03 & 0.26 $\pm$ 0.04 & 0.12 $\pm$ 0.02 & 0.15 $\pm $ 0.05 \\
 [0.2ex]
Attenuation $\lambda$ (fm) & 2.9  $\pm$ 0.3 & 3.7  $\pm$ 0.4 & 3.8  $\pm$ 0.6  & 8.1  $\pm$ 1.5 & 6.5 $\pm$ 2.0 \\ [0.2ex]
\hline
\hline
\end{tabular}
\caption{
\label{table:opacity}
Examples of opacities  $\kappa $ 
and attenuation lengths $\lambda = 1/\kappa $
in $\sqrt{s_{NN}}= 200$ GeV Au+Au reactions, 
evaluated from nuclear modification factors measured at $p_t = 4.75 $ GeV/c 
in ref.~\cite{Adare:2008qa} and HBT radii measured in ref.~\cite{Adler:2004rq},  
averaged over all directions  and charge combinations,
measured in the same centrality class as  $R_{AA}$. 
}
\end{table}

\section{Cross-checks}
In Table 1, optical opacities and nuclear attenuation lengths were calculated
from measured centrality dependent nuclear modification factors and HBT radius
parameters. In this section, some cross-checks are presented.

\begin{figure}[tbp]
\begin{center}
\includegraphics[width=0.7\textwidth]{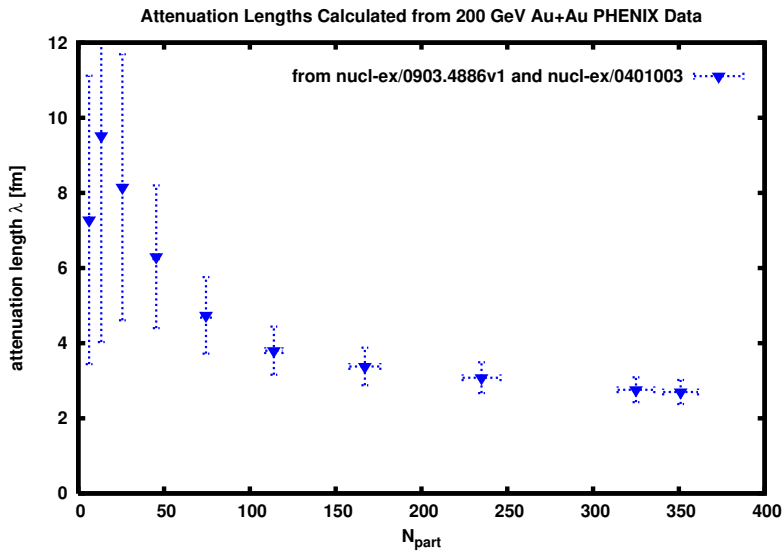}
\end{center}
\vspace{-0.5cm}
\caption{
Attenuation length in 200 GeV Au+Au collisions as a function of number of
participants, evaluated from PHENIX transverse momentum integrated 
nuclear modification factor $R_{AA} (p_t > 5 GeV)$ and number of participant
dependent HBT radius, $R_{HBT}$ measurements, 
using the formula $R_{HBT} = 0.83 + 0.54 N_{part}^{1/3}$.
}
\label{fig:lambda-Npart}
\end{figure}

A simple cross-check was to use the measured  nuclear modification
factor  of the transverse momentum integrated $R_{AA}(p_t > 5 GeV)$
from ref. ~\cite{Afanasiev:2009iv} 
and the HBT radii $R_{HBT}$
from ref. ~\cite{Adler:2004rq}, 
both measured  as a function of the number of participants.
In this case, the dependence of the HBT radii on the number of
participant was parameterized with a linear dependence in ref.
~\cite{Adler:2004rq} as $R_{HBT} = p_0 + p_1 N_{part}^{1/3}$. 
Using eq. (\ref{e:lambda}), the nuclear attenuation lengths in 200 GeV Au+Au collisions
were evaluated and plotted  as a function of number of
participants in Fig.~\ref{fig:lambda-Npart}.
This figure is consistent with Table~\ref{table:opacity}, that was obtained
from direct measurements of HBT radii and nuclear modification factors at a fixed
$p_t = 4.75$ GeV.  Fig.~\ref{fig:lambda-Npart} also indicates that
the nuclear attenuation length is the smallest in most central collisions,
and its range in these collisions is about 3 to 9 fm, depending on the centrality.

\begin{figure}[tbp]
\begin{center}
\includegraphics[width=0.7\textwidth]{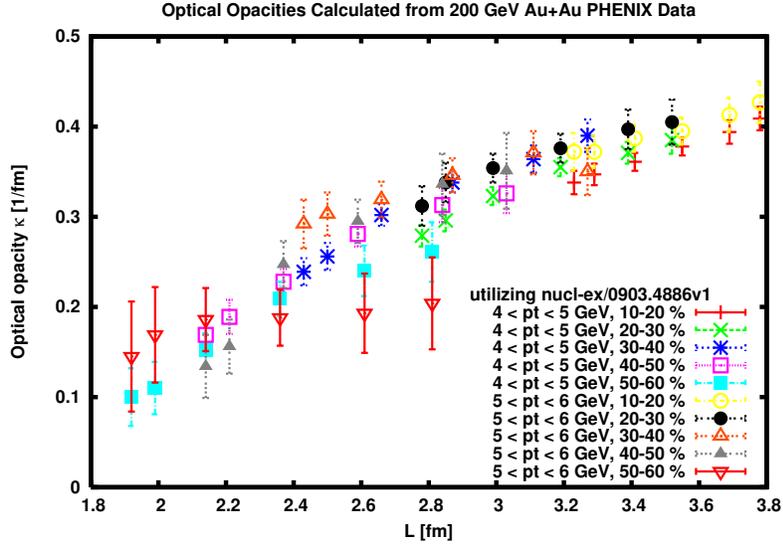}
\end{center}
\vspace{-0.5cm}
\caption{
Optical opacity, evaluated from PHENIX reaction plane angle dependent
nuclear suppression factor measurements and Glauber calculations of the
relevant length scale $L$.
}
\label{fig:kappa-L}
\end{figure}

Optical opacity can also be evaluated from the reaction plane angle dependent
nuclear suppression factor measurements of PHENIX
~\cite{Afanasiev:2009iv}   and the corresponding Glauber calculations of the
relevant length scale $L$, that corresponds to the lengthscale of the nuclear
geometry in the given direction.
Fig.~\ref{fig:kappa-L} 
indicates an approximate scaling law: if the nuclear
attenuation lengthscales are similar, but transverse momenta or centralities
are different, the measured optical opacities $\kappa$ are within errors
similar. This also indicates, that opacities characterize a local property
of the hot and dense matter. It is interesting to contrast this scaling law
to the $L$ dependence of the nuclear modification factors (Fig. 22 of
ref. ~\cite{Afanasiev:2009iv}), that change with centrality and 
transverse momentum even if the initial nuclear geometry or $L$ is similar.

\begin{figure}[tbp]
\begin{center}
\includegraphics[width=0.7\textwidth]{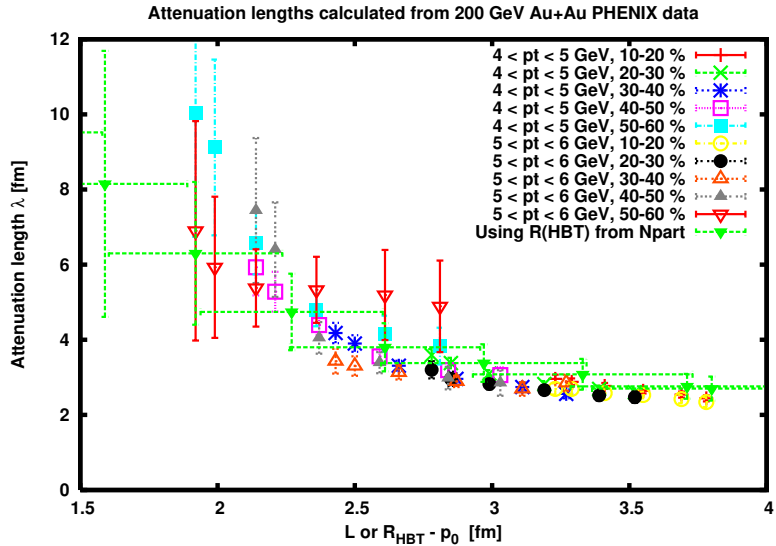}
\end{center}
\vspace{-0.5cm}
\caption{
Attenuation length, evaluated from PHENIX reaction plane angle dependent
nuclear suppression factor measurements and Glauber calculations and compared
with attenuation lenght evaluated from PHENIX pt integrated $R_{AA}$ measurement
combined with lenght-scale estimates from the centrality dependence of PHENIX HBT data,
$R_{HBT} = p_0 + p_1 N_{part}^{(1/3)}$.
Note that the two scaling laws coincide if the latter is plotted as a function of
$R_{HBT} - p_0$. 
}
\label{fig:kappa-HBT}
\end{figure}

It is also interesting to compare the results of 
Fig.~\ref{fig:lambda-Npart} and Fig~\ref{fig:kappa-L}.
The nuclear attenuation length, evaluated from PHENIX reaction plane angle dependent
nuclear suppression factor measurements and Glauber calculations is compared
with attenuation lenght evaluated from PHENIX $p_t$ integrated $R_{AA}$ measurement
combined with lenght-scale estimates from the centrality dependence of PHENIX HBT data,
$R_{HBT} = p_0 + p_1 N_{part}^{(1/3)}$.
Note that the two scaling laws coincide if the nuclear attenuation lenght 
is plotted as a function of $R_{HBT} - p_0$ or as a function of $L$,
as indicated in Fig.~\ref{fig:kappa-HBT}.

\section{Summary and Outlook}
I argued that for a definitive experimental program to
discover a QCD critical point, it is not sufficient to locate  only such a point on the baryochemical potential vs temperature or
 $(\mu_B,T)$ plane, but we also should measure at least two of its critical
exponents, and determine the universality class of such a CEP.

Even if we do not find a such a CEP,
but instead we can find a line of first order phase transitions using strong interactins,
I believe that the experimental program should be quantitative, and determine the speed of sound, 
the latent heat, the critical temperature and pressure of such a phase transition.

Note that this program is not a mission impossible, 
although some authors expressed  an opposite opinion
-- sometimes in no uncertain terms. 
In his CPOD  2009 talk, Rajagopal suggested, that critical exponents will
not/need not be measured in high energy heavy ion physics~\cite{Rajagopal:2009xx}:
{\it ``Since the correlation length cannot be larger than 2-3 fm, heavy ion experiments
can never be used to measure the critical exponents of the 2nd order critical point.
That's OK: we know that (its univerality class) is (that of the 3d) Ising (model).
What we don't know and we need experiments for, is \underline{where} it is located."}
With all due respect to Rajagopal, I disagree with his opinion,
and would like to encourage my experimentalist colleagues to determine the critical
exponents as well as the universality class of the QCD CEP, because  without the
measurement of these quantities a definitive search for the critical points 
cannot be realized, and the results cannot be compared with similar measurements
of other, well understood critical points, e.g. that of water. 
My arguments for experimental determination of the propertied of 
the QCD CEP are based on the following observations:
\begin{enumerate}

\item{} 
	In physics, theoretical predictions are
	compared to experimental results, as a matter of basic principle. 
	Only from a comparision with data it becomes 
	possible to select the best of competing  models/theories, 
	and new predictions are tested against new measurements to refine them further. 
	The existence of theoretical predictions in physics does not imply the lack of need
	for an experimental determination of key physical quantities,
	for example, the value of the critical exponents of the Critical Point of QCD.

\item{} In the theory of critical phenomena,
	there are static and dynamic universality classes, 
	and models that belong to
	the same static universality class may belong to different dynamical 
	universality classes~\cite{Hohenberg:1977ym}.
	For QCD, the static universality class is predicted to be 
	that of the 3d Ising model~\cite{Rajagopal:1992qz}, while the
	dynamic universality class is predicted to be different, corresponding to that of the 
	liquid-gas phase transition~\cite{Son:2004iv}.

\item{} In a heavy ion collisions, violent initial dynamics may create random
fields that in fact may modify the critical dynamics as well. The presence of
random fields is known to modify the universality class and the critical exponents of
the 3d Ising model, for example the critical exponent of the correlation function increases dramatically
from the value of $0.03 \pm 0.01$ to $0.50 \pm 0.05$ 
if the universality class of the 3d Ising model is changed to that of the random field
3d Ising model, see e.g. ref.~\cite{Csorgo:2009gb} for more details.

\item{} It has been demonstrated experimentally that another critical phenomena,
namely nuclear multifragmentation can be studied in p+Xe collisions at $ 1 \le E_p \le 20$ GeV
measuring the inclusive production of light fragments $(3 \le Z_f \le 17)$, ref.~\cite{Mahi:1988hj}.
From a measurement of intermediate mass fragment production
in heavy ion collisions with 1 A GeV beam energies  of Au, La, and Kr,
the critical exponents of the nuclear liquid-gas phase transition were
 extracted at these relatively low energies. The results allowed
for a discrimination among various universality classes, see Table I 
of ref.~\cite{Srivastava:2002xx}. If the universality class and the critical exponents
can be determined reasonably well in these low multiplicity heavy ion and proton induced reactions 
at relatively moderate energies,
extracted from distributions that have less than 2 orders of 
magnitude vertical range, then even better measurements should be possible in the RHIC
low energy scan and in the SPS future heavy ion program, because
these measurements span already several orders of magnitude vertical scales e.g. in the
transverse momentum distributions, and simply because the number of produced particles
is substantially larger than that of the nuclear fragments.

\item{}
	I would also like to draw attention to the presentation of J.T. Mitchell~\cite{Mitchell:CPOD09}
where he reported on a search for a QCD Critical Point at RHIC
using various observables that were already published or are in 
a preliminary data status, which include PHENIX preliminary results for
observables that become critical exponents at the CEP, but of course
they can be measured at other points of the phase diagram, too.
Using the presently available PHENIX/RHIC data set, no significant indication
for the existence of a QCD critical point was seen. Further searches are in progress 
in STAR and PHENIX, in the forthcoming RHIC low energy scan program.
\end{enumerate}

Based on these arguments, the experimental identification, location, characterization
and cross-checking of the properties of a QCD critical point is apparently 
experimentally possible and scientifically desirable.

\section{Summary and conclusions}
Critical opalescence is a smoking gun signature of a Critical Point.  In QCD,
optical opacity is a function of nuclear modification factor $R_{AA}$ and the distance
covered by the attenuated jet in the medium, which is proportional to the initial
nuclear geometry, for example $N_{part}^{1/3}$. A lengthscale with such a known centrality
dependence is measured by femtoscopy, with HBT correlation techniques, called $R_{HBT}$. The combination of
the nuclear modification factor with such a length scale can be used to measure
optical opacity $\kappa$ or nuclear attenuation length $\lambda$.
Azimuthally sensitive nuclear modification factor and  HBT radius 
measurements allow directional dependent studies of spatial dimensions, 
that add valuable information to directional dependent
$R_{AA}$ measurements. 

By measuring the critical exponents of the correlation 
function and the correlation length with methods discussed below, 
a first estimate of the universality class of the QCD critical point can be given.

\begin{enumerate}
\item{}
Search for critical opalescence,
i.e. looking  for the maximum of opacity, or the minimum of attenuation length,
as a function of centrality, bombarding energy and size of the colliding nuclei in a broad,
high transverse momentum range.
\item{}
Search for a non-monotonous behaviour of the correlation length as a function
of colliding energy and centrality (related to critical exponent $\alpha$),
from multiplicity fluctuation measurement and fits of Negative Binomial Distributions
to multiplicity distributions in limited rapidity intervals.
This method can locate experimentally the critical end point (CEP) of the line of 
1st order phase transitions, if such a CEP exists.
\item{}
Search a minimum of the exponent of the correlation  function, $\eta$,
as a function of colliding energy and centrality, using L\'evy fits to the Bose-Einstein (HBT)
correlation function.
\item{}
 Determine the chemical freeze-out temperature $T$ on an event-by-event basis
for an event selection that corresponds to the maximum of opacity.
Check that various observables (heat capacity, correlation length) indeed
indicate a power-law behavior as a function of $t = |T-T_c|/T_c$ simultaneously.
\item{}
The correlation length can be
determined from the multiplicity distribution using the negative binomial distribution, and
the exponent of the correlation length $\nu$ can be obtained from event-by-event determination
of {\it both} the correlation length and the chemical freeze-out temperature $T$.
\item{}
Determine the critical exponents for at least 
two observables, recommended exponents are $\eta$ and $\nu$. 
Use theoretical relations
to determine the remaining critical exponents, so that all the six critical exponents be given,
providing the universality class of the QCD critical point.
The critical exponent of the correlation function $\eta$ 
is defined in the momentum-space, and the power-law shape of the correlation function
shows up as a function of the relative momentum of particle 
pairs and not as a function of the reduced temperature $t$, which requires an event-by-event analysis.
Hence this  critical exponent  $\eta$ is relatively easy to measure. 
\item{}
Cross-check the result by measuring additional 
critical exponents and cross-check these measurements with constraints 
obtained from the theory of second order phase transitions.
For example, determining the exponent of the heat capacity $\alpha$ 
seems to be straight-forward, as the heat capacity is measured by
the event-by-event fluctuations of the chemical freeze-out temperature. 
\end{enumerate}

Although it is an extremely challenging task, that some may even consider as
mission impossible, the above items indicate that we realistically may 
locate the critical point and also to estimate at least two of the critical exponents, hence
the universality class of the QCD Critical End Point. These measurements will be possible
in the low energy scan program at  RHIC, in a future CERN SPS heavy ion program and also utilizing 
the upcoming FAIR facility.

\acknowledgments
I would like to thank Frithiof Karsch, Paul Sorensen and the Local Organizing Committee of CPOD 2009 for creating
an inspiring and useful workshop atmosphere. It is my pleasure to thank professor Roy Glauber for his
kind hospitality at Harvard University.

\vfill\eject

\end{document}